\documentclass[aps,prd,preprint,groupedaddress,showpacs]{revtex4}
\usepackage[dvips]{graphicx}
\begin{document}
\preprint{KOBE-TH-03-01}
\title{Stability of Neutral Fermi Balls with Multi-Flavor Fermions}

\author{T.Yoshida}
\affiliation{Department of Physics, Tokyo University, Hongo 7-3-1, Bunkyo-Ku, Tokyo 113-0033, Japan}

\author{K.Ogure}
\affiliation{Department of Physics, Kobe University, Rokkoudaicho 1-1, Nada-Ku, Kobe 657-8501, Japan}

\author{J.Arafune}
\affiliation{National Institution for Academic Degrees, Hitotsubashi 2-1-2, Chiyoda-Ku, Tokyo 101-8438, Japan}

\date{\today}

\begin{abstract}
A Fermi ball is a kind of non-topological soliton, which is thought to
arise from the spontaneous breaking of an approximate $Z_2$ symmetry and
to contribute to cold dark matter. We consider a simple model in which
fermion fields with multi-flavors are coupled to a scalar field through
Yukawa coupling, and examine how the number of the fermion flavors
affects the stability of the Fermi ball against the fragmentation. (1)We
find that the Fermi ball is stable against the fragmentation in most
cases even in the lowest order thin-wall approximation. (2)We then find
that in the other specific cases, the stability is marginal in the
lowest order thin-wall approximation, and the next-to-leading order
correction determines the stable region of the coupling constants; We
examine the simplest case where the total fermion number $N_i$ and the
Yukawa coupling constant $G_i$ of each flavor $i$ are common to the
flavor, and find that the Fermi ball is stable in the limited region of
the parameters and has the broader region for the larger number of the
flavors.
\end{abstract}

\maketitle

%%%%%%%%%%
%%%%%%%%%%
\section{INTRODUCTION}
%%%%%%%%%%
%%%%%%%%%%
\label{intro.sec}

A Fermi ball \cite{MAC,MOR}, a kind of non-topological soliton
\cite{LEE}, is composed of three parts: a false vacuum domain, a domain
wall enveloping the domain, and zero-mode fermions \cite{DVA} confined
in the domain wall. The Fermi ball is stabilized owing to the dynamical
balance of the shrinking force due to the surface energy and the volume
energy, and the expanding force due to the Fermi energy. The Fermi ball
is thought to be a candidate for one kind of cold dark matter in the
present universe \cite{ARA}.

Macpherson and Campbell pointed out that such stability holds good only
for the spherical shape of the Fermi ball \cite{MAC}. They further
showed that the Fermi ball is not stable against the deformation of the
spherical shape, and thus flattens and fragments into tiny Fermi
balls. The destabilization is caused by the volume energy of the Fermi
ball.

We, however, pointed out that the perturbative correction due to the
domain wall curvature can stabilize the Fermi ball when the volume
energy is small enough compared to the curvature effect
\cite{YOS,OGU}. In case of a simple model with a single fermion flavor,
we found that only in the quite narrow region of the parameters does the
Fermi ball become stable.

The purpose of the present paper is to examine how the fermion content
of the model affects the stability of the Fermi ball. As an example, we
consider an extended model in which fermions with multi-flavors are
coupled to a scalar field through Yukawa coupling. Since the Pauli's
exclusion principle does not apply to the different flavors of the
fermions, the stable region of the parameters is expected to broaden.

%%%%%%%%%%
%%%%%%%%%%
\section{STABILITY OF FERMI BALL}
%%%%%%%%%%
%%%%%%%%%%\label{stable.sec}
%%%%%%%%%%
%%%%%%%%%%
We consider the following Lagrangian density, 
\begin{equation}
{\cal L}=\frac{1}{2}(\partial_{\mu}\phi)^2 
	+\sum_{i=1}^{n}\overline{\Psi_i}
	(i\gamma_i^{\mu}\partial_{\mu}-G_i\phi)\Psi_i -U(\phi)~~,
\label{L.eqn2}
\end{equation}
where the scalar potential $U(\phi)$ is given by
\begin{equation}
U(\phi)=\frac{\lambda}{8}(\phi^2-v^2)^2 +\Delta(\phi) ~~.
\label{U.eqn2}
\end{equation}
If the quantity $|\Delta(v)-\Delta(-v)|$ is zero, the Lagrangian density
is invariant under the $Z_2$ transformation, $\phi \leftrightarrow
-\phi$.  There is, however, a small but a finite quantity
$|\Delta(v)-\Delta(-v)|\simeq \Lambda \ll \lambda v^4$, where the
invariance is not a strict one.

We consider a spherical Fermi ball with the radius $R$, and assume
that the wave function $\Psi_i$ and the boson $\phi$ are static and that
$\phi$ depends only on the radial coordinate $r$. Let $\Psi_i$ be the
eigenfunction of the total angular momentum squared ${\vec {\bf J}}^2$,
the $z$ component ${\bf J_z}$ and the parity ${\bf P}$ with the
eigenvalues of $J(J+1)$, $M$ and $(-1)^{J-\omega/2}$ ($\omega=\pm 1$),
respectively. Then, $\Psi_i$ is written as
\begin{eqnarray}
\Psi_i(\vec{x})=\frac{1}{r} \left(
	\begin{array}{l}
	f(r){\cal Y}_{lJ}^{M}(\theta,\varphi) \\
	g(r){\cal Y}_{l'J}^{M}(\theta,\varphi)
	\end{array}
\right)~~,
\label{fYgY.eqn2}
\end{eqnarray}
where ${\cal Y}_{lJ}^{M}$ and ${\cal Y}_{l'J}^{M}=({\vec \sigma}{\vec x}/r){\cal Y}_{lJ}^{M}$ are the spherical spinors having the eigenvalues $J$ and $M$, with $J=l+\omega/2=l'-\omega/2$.  Substituting Eq.(\ref{fYgY.eqn2}) into the Lagrangian $L=\int {\rm d}^3x~{\cal L}$, we obtain
\begin{equation}
L[\phi,\psi_i]=
	-\int_{0}^{\infty}{\rm d}r~
	\left[
	4\pi r^2\left\{
		\frac{1}{2}\left(\frac{{\rm d}\phi}{{\rm d}r}\right)^2
		+U(\phi)\right\}
	+\sum_i\sum_{KM}
	\psi_i^{\dagger}H_f\psi_i
	\right]~~,
\label{LL.eqn2}
\end{equation}
where
\begin{equation}
H_f=\sigma_1\frac{1}{i}\frac{{\rm d}}{{\rm d}r}
	+\sigma_2\frac{K}{r}
	+\sigma_3 G_i\phi~~,
\label{Hf.eqn2}
\end{equation}
with $K=\omega (J+\frac{1}{2})$ and $\psi_i(r)={f(r)\choose g(r)}$. Since the Fermi ball is a ground state with a fixed number of fermions,
\begin{equation}
N_i=\int{\rm d}^3x~\Psi_i^{\dagger}\Psi_i~~,
\label{Nf.eqn2}
\end{equation}
we obtain the wave function $\psi_i$ and the scalar field $\phi$ by extremizing
\begin{equation}
L_{\epsilon}[\phi,\psi_i]=L[\phi,\psi_i]
	+\sum_i\epsilon_i \left(
		\sum_{KM}\int_{0}^{\infty}{\rm d}r~
			\psi_i^{\dagger}\psi_i -N_i\right)~~,
\label{Le.eqn2}
\end{equation}
with the Lagrange multipliers $\epsilon_i$. The energy of the Fermi ball is expressed in terms of the fields as
\begin{equation}
E=\int_{0}^{\infty}{\rm d}r~
	\left[ 4\pi r^2\left\{
		\frac{1}{2}\left( \frac{{\rm d}\phi}{{\rm d}r} \right)^2 
		+U(\phi) \right\}\right]
	+\sum_i\sum_{KM}\epsilon_i
\label{E.eqn2}
\end{equation}
where $\epsilon_i$ is equal to the Fermi energy $\epsilon_i=\int_0^{\infty}{\rm d}r~\psi_i^{\dagger}H_f\psi_i$ and $\psi_i$ is normalized as $\int_0^{\infty}{\rm d}r~\psi_i^{\dagger}\psi_i = 1$.  In order to estimate the energy of the Fermi ball, we take the thin-wall approximation and obtain the correction due to the finite curvature radius $R$ by the perturbation with respect to $1/R$. We expand $\phi$, $\psi_i$, and $H_f$ in the power of $1/R$,
\begin{eqnarray}
\left\{
\begin{array}{l}
\phi = \phi_0 + \phi_1 +\cdots \\
\psi_i = \psi_{i0} +\psi_{i1} +\cdots \\
H_f = H_0 + H_1 + H_2 +\cdots~~,
\end{array}
\right.
\label{expand.eqn2}
\end{eqnarray}
where
\begin{eqnarray}
H_0 &=& 
	\sigma_1\frac{1}{i}\frac{{\rm d}}{{\rm d}r}
	+\sigma_2\frac{K}{R}
	+\sigma_3 G_i\phi_0 
\nonumber \\
H_1 &=&
	-\sigma_2 \frac{Kw}{R^2}
	+\sigma_3 G_i\phi_1 
\nonumber \\
H_2 &=&
	\sigma_2 \frac{Kw^2}{R^3}~~,
\label{H0H1H2.eqn2}
\end{eqnarray}
with $w=r-R$. From $\delta L_{\epsilon}/\delta \phi = \delta L_{\epsilon}/\delta \psi_i^{\dagger} =0$, we obtain the equations of motion,
\begin{eqnarray}
&&H_0\psi_{i0}=\epsilon_{i0}\psi_{i0}
\label{eqM0psi.eqn2} \\
&&\frac{{\rm d}^2\phi_0}{{\rm d}w^2}
=\left. \frac{\partial U}{\partial \phi} \right|_{\phi=\phi_0}
+\sum_i\frac{G_i}{4\pi R^2}\sum_{KM}\psi_{i0}^{\dagger}\sigma_3\psi_{i0}~~,
\label{eqM0phi.eqn2}
\end{eqnarray}
and
\begin{eqnarray}
&&(H_0-\epsilon_{i0})\psi_{i1}=-(H_1-\epsilon_{i1})\psi_{i0}
\label{eqM1psi.eqn2} \\
&&\left[
	\frac{{\rm d}^2}{{\rm d}w^2}
	-\left. \frac{\partial^2 U}{\partial \phi^2} \right|_{\phi=\phi_0}
\right]\phi_1
=
-\frac{2}{R}\frac{{\rm d}\phi_0}{{\rm d}w}
+\sum_i\frac{G_i}{2\pi R^2}\sum_{KM}\psi_{i0}\sigma_3\psi_{i1}~~.
\label{eqM1phi.eqn2}
\end{eqnarray}
Neglecting $\Delta (\phi)$ in the scalar potential $U(\phi)$ for simplicity, we have analytic solutions for $\phi_0$ and $\psi_{i0}$,
\begin{eqnarray}
&&\phi_0(w)=
	v\tanh{\frac{w}{\delta_b}} 
\label{phi0.eqn2} \\
&&\psi_{i0}(w)=
	\frac{1}{\sqrt{{\cal N}_i}}\frac{1}{{\rm cosh}^{\gamma_i}\frac{w}{\delta_b}} \chi_+~~,
\label{psi0.eqn2}
\end{eqnarray}
where $\delta_b=2\lambda^{-\frac{1}{2}}v^{-1}$ is the thickness of the domain wall, $\gamma_i = 2\lambda^{-\frac{1}{2}}G_i$ is the constant\footnote{The constant $\gamma_i$ is equal to the squared ratio of the thickness of the domain wall to that of the distribution of the fermion confined in the wall, $\gamma_i = (\delta_b/\delta_f)^2$, where $\delta_f$ is given by $\delta_f=\sqrt{2}\lambda^{-\frac{1}{4}}G_i^{-\frac{1}{2}}v^{-1}$.}, ${\cal N}_i=\int_{-\infty}^{+\infty}{\rm d}w~{\rm cosh}^{-2\gamma_i}w/\delta_b$ is the normalization factor, and $\chi_{\pm}$ is the eigenspinor of $\sigma_2$ with the eigenvalue $\pm 1$. We note that the second term of the r.h.s. of Eq.(\ref{eqM0phi.eqn2}) vanishes. The leading order of the eigenvalue is given by
\begin{equation}
\epsilon_{i0}=\frac{K}{R}~~,
\label{epsilon0.eqn2}
\end{equation}
where we take $K$ positive ($\omega = +1$). We have solutions for $\phi_1$ and $\psi_{i1}$,
\begin{eqnarray}
&&\phi_1(w)=
	\frac{1}{{\rm cosh}^2\frac{w}{\delta_b}}
	\int_{0}^{w}{\rm d}w'~ {\rm cosh}^4\frac{w'}{\delta_b}
	\int_{0}^{w'}{\rm d}w''~ 
		\frac{h(w'')}{{\rm cosh}^2\frac{w''}{\delta_b}} 
\label{phi1.eqn2} \\
&&\psi_{i1}(w)=
	\frac{1}{\sqrt{{\cal N}_i}}
	\biggl\{
	c_{i+}(w)\chi_{+}+c_{i-}(w)\chi_{-}
	\biggr\}~~,
\label{psi1.eqn2}
\end{eqnarray}
where
\begin{equation}
h(w)=-\frac{2v}{\delta_b R{\rm cosh}^2\frac{w}{\delta_b}}
	+\frac{1}{2\pi R^4}\sum_i\sum_{KM}
	\frac{KG_i}{{\cal N}_i}\int_w^{\infty}{\rm d}w'~
		\frac{w'}{{\rm cosh}^{2\gamma_i}\frac{w'}{\delta_b}}~~,
\label{h.eqn2}
\end{equation}
and 
\begin{eqnarray}
&&c_{i+}(w)= \frac{1}{{\rm cosh}^{\gamma_i}\frac{w}{\delta_b}}
	\Biggl\{
	\frac{2K^2}{R^3}\int_0^w{\rm d}w'~ {\rm cosh}^{2\gamma_i}\frac{w'}{\delta_b}
	\int_{w'}^{\infty}{\rm d}w''~ 
	\frac{w''}{{\rm cosh}^{2\gamma_i}\frac{w''}{\delta_b}}
	\Biggr. 
\nonumber \\
&&\hspace{3cm}
	\Biggl.
	-G_i\int_{0}^{w}{\rm d}w'~ \phi_1(w')
	\Biggr\} 
\nonumber \\
&&c_{i-}(w)= \frac{K}{R^2}{\rm cosh}^{\gamma_i}\frac{w}{\delta_b}
	\int_w^{+\infty}{\rm d}w'~
		\frac{w'}{{\rm cosh}^{2\gamma_i}\frac{w'}{\delta_b}}~~.
\label{cpm.eqn2}
\end{eqnarray}
Substituting the solutions into Eq.(\ref{E.eqn2}), we obtain the energy of the Fermi ball,
\begin{equation}
E= E_0+\delta E~~,
\label{EE.eqn2}
\end{equation}
where $E_0$ is the leading order contribution to the energy,
\begin{equation}
E_0=\frac{8\pi \lambda^{\frac{1}{2}}v^3R^2}{3}
	+\frac{2\sum_iN_i^{\frac{3}{2}}}{3R}~~,
\label{E0.eqn2}
\end{equation}
and $\delta E$ is the energy correction of the order of $E_0\times(\delta_b/R)^2$,
\begin{eqnarray}
\delta E&=& \frac{\sum_iN_i^{\frac{1}{2}}}{12R}
	+\pi \lambda v^4\int_{-\infty}^{+\infty}{\rm d}w~
		\frac{w^2}{{\rm cosh}^4\frac{w}{\delta_b}}
	\nonumber \\
&&	-2\pi \lambda^{\frac{1}{2}}v^2R\int_{-\infty}^{+\infty}{\rm d}w~
		\frac{1}{{\rm cosh}^4\frac{w}{\delta_b}}
		\int_{0}^{w}{\rm d}w'~{\rm cosh}^4\frac{w'}{\delta_b}
		\int_{0}^{w'}{\rm d}w''~
			\frac{h(w'')}{{\rm cosh}^2\frac{w}{\delta_b}}
	\nonumber \\
&&	+\frac{2}{3R^3}\sum_i\frac{N_i^{\frac{3}{2}}}{{\cal N}_i}
	\int_{-\infty}^{+\infty}{\rm d}w~
		\frac{w^2}{{\rm cosh}^{2\gamma_i}\frac{w}{\delta_b}}
	\nonumber \\
&&	-\frac{4}{5R^5}\sum_i\frac{N_i^{\frac{5}{2}}}{{\cal N}_i}
	\int_{-\infty}^{+\infty}{\rm d}w~
		\frac{w^2}{{\rm cosh}^{2\gamma_i}\frac{w}{\delta_b}}
	\int_{0}^{w}{\rm d}w'~{\rm cosh}^{2\gamma_i}\frac{w'}{\delta_b}
	\nonumber \\
&& \hspace{4cm}
	\times
	\int_{w'}^{\infty}{\rm d}w''~
		\frac{w''}{{\rm cosh}^{2\gamma_i}\frac{w''}{\delta_b}}
	\nonumber \\
&&	+\frac{2}{3R^2}\sum_i\frac{G_iN_i^{\frac{3}{2}}}{{\cal N}_i}
	\int_{-\infty}^{+\infty}{\rm d}w~
		\frac{w}{{\rm cosh}^{2\gamma_i}\frac{w}{\delta_b}}
	\int_{0}^{w}{\rm d}w'~
		\frac{1}{{\rm cosh}^2\frac{w'}{\delta_b}}
	\nonumber \\
&& \hspace{2cm}
	\times
	\int_{0}^{w'}{\rm d}w''~{\rm cosh}^4\frac{w''}{\delta_b}
	\int_{0}^{w''}{\rm d}w'''~
		\frac{h(w''')}{{\rm cosh}^2\frac{w''}{\delta_b}}~~.
	\label{deltaE.eqn2}
\end{eqnarray}
In the above equations, we use the relations,
\begin{eqnarray}
N_i&=&\sum_{KM}=\sum_{K=1}^{K_{max}}\sum_{M=-J}^{J}
	=\sum_{K=1}^{K_{max}}(2K)
	=K_{max}(K_{max}+1)~~,
\label{Kmax.eqn2}\\
\sum_{KM}\epsilon_{i0}&=&\frac{1}{R}\sum_{KM}K
	=\frac{2}{3R}K_{max}\bigl(K_{max}+1\bigr)
		\bigl(K_{max}+\frac{1}{2}\bigr)
\nonumber \\
&\simeq & \frac{2N_i^{\frac{3}{2}}}{3R}+\frac{N_i^{\frac{1}{2}}}{12R}
\hspace{1.5cm} (N_i \gg 1)~.
\label{fermienergy0.eqn2}
\end{eqnarray}

(1){\it Stability in the leading order approximation}\\
Let us examine the stability of the Fermi ball within the leading order approximation in the $\delta_b/R$-expansion.  From $\partial E_0/\partial R =0$, we get the minimizing radius,
\begin{equation}
R_{min}=\frac{\left(\sum_iN_i^{\frac{3}{2}}\right)^{\frac{1}{3}}}{2\pi^{\frac{1}{3}}\lambda^{\frac{1}{6}}v}~,
\label{R.eqn2}
\end{equation}
and the energy at the radius,
\begin{equation}
E_0=2\pi^{\frac{1}{3}}\lambda^{\frac{1}{6}}
	\biggl(\sum_iN_i^{\frac{3}{2}}\biggr)^{\frac{2}{3}}v~~.
\label{E00.eqn2}
\end{equation}
we note $\partial^2 E_0/\partial R^2 > 0$ at $R=R_{min}$.

In order to examine the stability against the fragmentation, we compare
two states; a state ${\cal A}$ in which a single Fermi ball has the
fermion number $N_i$ for $i$-th flavor, and a state ${\cal B}$ in which
$m$ Fermi balls have the fermion number $N_i^{(a)}$ each and conserve
the total fermion number as $\sum_{a=1}^{m}N_i^{(a)}=N_i$ for each
flavor. States ${\cal A}$ and ${\cal B}$ have the energy $E_{\cal
A}=E_0(N_i)$ and $E_{\cal B}=\sum_aE_0(N_i^{(a)})$, respectively. To
compare the energy of the two states, we use Minkowski's inequality,
\begin{equation}
\biggl(\sum_i(N_i^{(1)}+N_i^{(2)})^{\frac{3}{2}}\biggr)^{\frac{2}{3}}
\leq
\biggl(\sum_i(N_i^{(1)})^{\frac{3}{2}}\biggr)^{\frac{2}{3}}
+\biggl(\sum_i(N_i^{(2)})^{\frac{3}{2}}\biggr)^{\frac{2}{3}}~,
\label{minkowski.eqn2}
\end{equation}
where the equality is valid only for $N_i^{(2)}=cN_i^{(1)}$ ($c\geq 0$) with $c$ being common for all $i$. Using the relation repeatedly, we have
\begin{equation}
\biggl(\sum_i(N_i)^{\frac{3}{2}}\biggr)^{\frac{2}{3}}
\leq
\sum_a\biggl(\sum_i(N_i^{(a)})^{\frac{3}{2}}\biggr)^{\frac{2}{3}}~,
\label{minkowskii.eqn2}
\end{equation}
where the r.h.s. is equal to the l.h.s. only for $N_i^{(a)}=c^{(a)}N_i$
($c^{(a)}\geq 0$ and $\sum_{a=1}^{m}c^{(a)}=1$). This leads us to the
fact that except for the special case of $N_i^{(a)}=c^{(a)}N_i$, state
${\cal A}$ has lower energy than that of state ${\cal B}$, and thus
the Fermi ball is stable against the fragmentation in the leading order
approximation. This situation that the Fermi ball is stable in most
cases is characteristic of the case with multi-flavor of fermions, and
qualitatively different from the case of a single flavor \cite{OGU}. In
case of $N_i^{(a)}=c^{(a)}N_i$, the two states have the same energy in
the leading order approximation, and the correction term $\delta E$
determines the stability of the Fermi ball against the fragmentation.

(2){\it Stability in the next-to-leading order approximation in the special case $N_i^{(a)}=c^{(a)}N_i$}\\
We examine the stability of the Fermi ball against the fragmentation in the case of $N_i^{(a)}=c^{(a)}N_i$. Substituting $R=R_{min}$ into Eq.(\ref{deltaE.eqn2}) yields
\begin{equation}
\delta E={\cal C}(\lambda,G_i,N_i)v~,
\label{deltaEE.eqn2}
\end{equation}
where
\begin{eqnarray}
&&
{\cal C}(\lambda,G_i,N_i)=
\frac{\pi^{\frac{1}{3}}\lambda^{\frac{1}{6}}\bigl(\sum_iN_i^{\frac{1}{2}}\bigr)}{6\bigl(\sum_iN_i^{\frac{3}{2}}\bigr)^{\frac{1}{3}}}
+\frac{8\pi(I_1-I_2)}{\lambda^{\frac{1}{2}}}
+\frac{64\pi\bigl(\sum_iN_i^{\frac{3}{2}}\overline{{\cal N}_i}I_3(i)\bigr)}{3\lambda^{\frac{1}{2}}\bigl(\sum_iN_i^{\frac{3}{2}}\bigr)}
\nonumber \\
&& \hspace{1.6cm}
-\frac{2048\pi^{\frac{5}{3}}\bigl(\sum_iN_i^{\frac{5}{2}}\overline{{\cal N}_i}I_4(i)\bigr)}{5\lambda^{\frac{7}{6}}\bigl(\sum_iN_i^{\frac{3}{2}}\bigr)^{\frac{5}{3}}}
+\frac{128\pi\bigl(\sum_iG_iN_i^{\frac{3}{2}}\overline{{\cal N}_i}I_5(i)\bigr)}{3\lambda \bigl(\sum_iN_i^{\frac{3}{2}}\bigr)}~.
\label{C.eqn2}
\end{eqnarray}
Here, $I_1$ to $I_5$ are given by
\begin{eqnarray}
&&I_1=
\int_{-\infty}^{+\infty}{\rm d}x~
\frac{x^2}{{\rm cosh}^4x} 
\nonumber \\
&&I_2=
\int_{-\infty}^{+\infty}{\rm d}x~
\frac{1}{{\rm cosh}^4x}
\int_{0}^{x}{\rm d}x'~
{\rm cosh}^4x'
\int_{0}^{x'}{\rm d}x''~
\frac{\bar{h}(x'')}{{\rm cosh}^2x''} 
\nonumber \\
&&I_3(i)=
\int_{-\infty}^{+\infty}{\rm d}x~
\frac{x^2}{{\rm cosh}^{2\gamma_i}x} 
\nonumber \\
&&I_4(i)=
\int_{-\infty}^{+\infty}{\rm d}x~
\frac{x}{{\rm cosh}^{2\gamma_i}x}
\int_{0}^{x}{\rm d}x'~
{\rm cosh}^{2\gamma_i}x'
\int_{x'}^{+\infty}{\rm d}x''~
\frac{x''}{{\rm cosh}^{2\gamma_i}x''} 
\nonumber \\
&&I_5(i)=
\int_{-\infty}^{+\infty}{\rm d}x~
\frac{x}{{\rm cosh}^{2\gamma_i}x}
\int_{0}^{x}{\rm d}x'~
\frac{1}{{\rm cosh}^2x'}
\int_{0}^{x'}{\rm d}x''~
{\rm cosh}^4x'' 
\nonumber \\
&&\hspace{6cm} \times
\int_{0}^{x''}{\rm d}x'''~
\frac{\bar{h}(x''')}{{\rm cosh}^2x'''}~,
\label{Is.eqn2}
\end{eqnarray}
with $h(w)$ rescaled as $\bar{h}(x)=\frac{R\delta_b}{v}h(\delta_bx)$ and
${\cal N}_i$ as $\overline{{\cal N}_i}=\frac{1}{\delta_b}{\cal N}_i$. We
compare state ${\cal A}$ of the single Fermi ball and state
${\cal B}$ of $m$ Fermi balls with the total fermion number to be
conserved for each flavor. States ${\cal A}$ and ${\cal B}$ have the
energy $E_{\cal A}=E_0(N_i)+{\cal C}(\lambda,G_i,N_i)v$ and $E_{\cal
B}=\sum_aE_0(N_i^{(a)})+\sum_a{\cal C}(\lambda,G_i,N_i^{(a)})v$,
respectively. In case of $N_i^{(a)}=c^{(a)}N_i$, we derive
$\sum_aE_0(N_i^{(a)})=E_0(N_i)$ from Eq.(\ref{E00.eqn2}) and ${\cal
C}(\lambda,G_i,N_i^{(a)})={\cal C}(\lambda,G_i,N_i)$ from
Eq.(\ref{C.eqn2}), and thus find that state ${\cal B}$ has the energy
$E_{\cal B}=E_0(N_i)+m{\cal C}(\lambda,G_i,N_i)v$. Therefore, if ${\cal
C}(\lambda,G_i,N_i)$ is positive, state ${\cal A}$ has lower energy than
that of the state ${\cal B}$ by the magnitude of the correction term
$\delta E$, and the Fermi ball is stable against fragmentation even
in the special case of $N_i^{(a)}=c^{(a)}N_i$.

\begin{figure}[htbp]
\centering
\includegraphics[height=9cm]{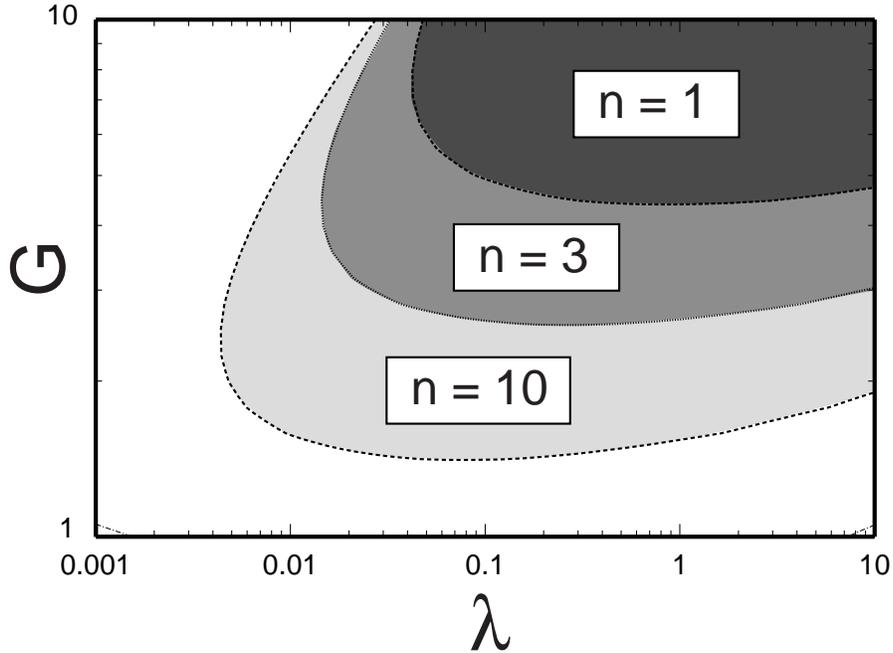}
\caption{The allowed regions (shadowed) of the scalar self-coupling constant $\lambda$ and the Yukawa coupling constant $G$ for the Fermi ball to be stable against the fragmentation. We assume that the fermion $\Psi_i$ ($1\leq i\leq n$) belongs to a multiplet and the boson $\phi$ to a singlet of the internal symmetry, and that the fermion number $N_i$ is common to the flavor as $N_i=N$. The figure shows that the allowed region broadens as $n$ increases. \label{onphase.fig}}
\end{figure}

\begin{figure}[htbp]
\begin{minipage}[b]{.5\linewidth}
\includegraphics[width=8cm]{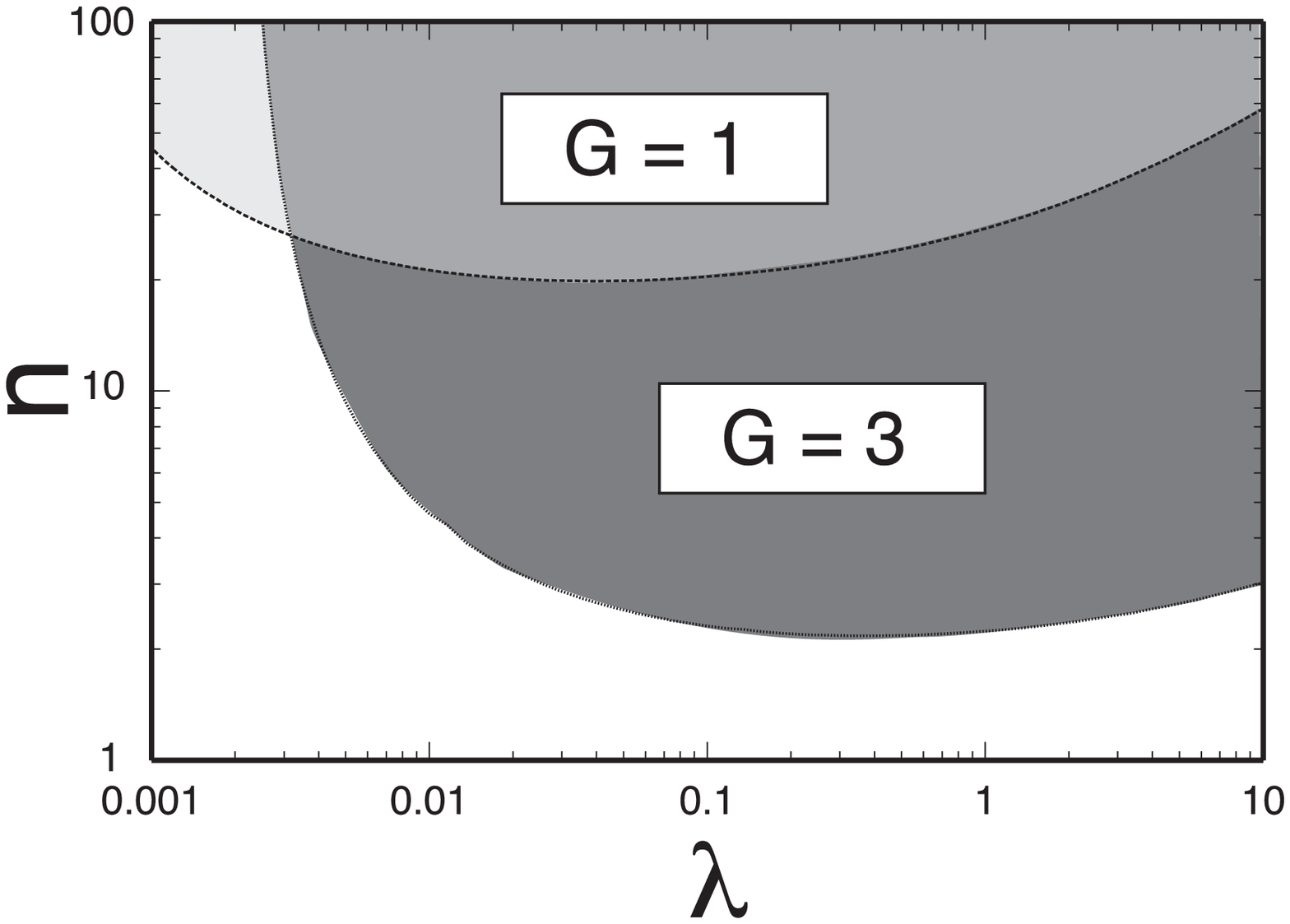}
\end{minipage}
\begin{minipage}[b]{.5\linewidth}
\includegraphics[width=8cm]{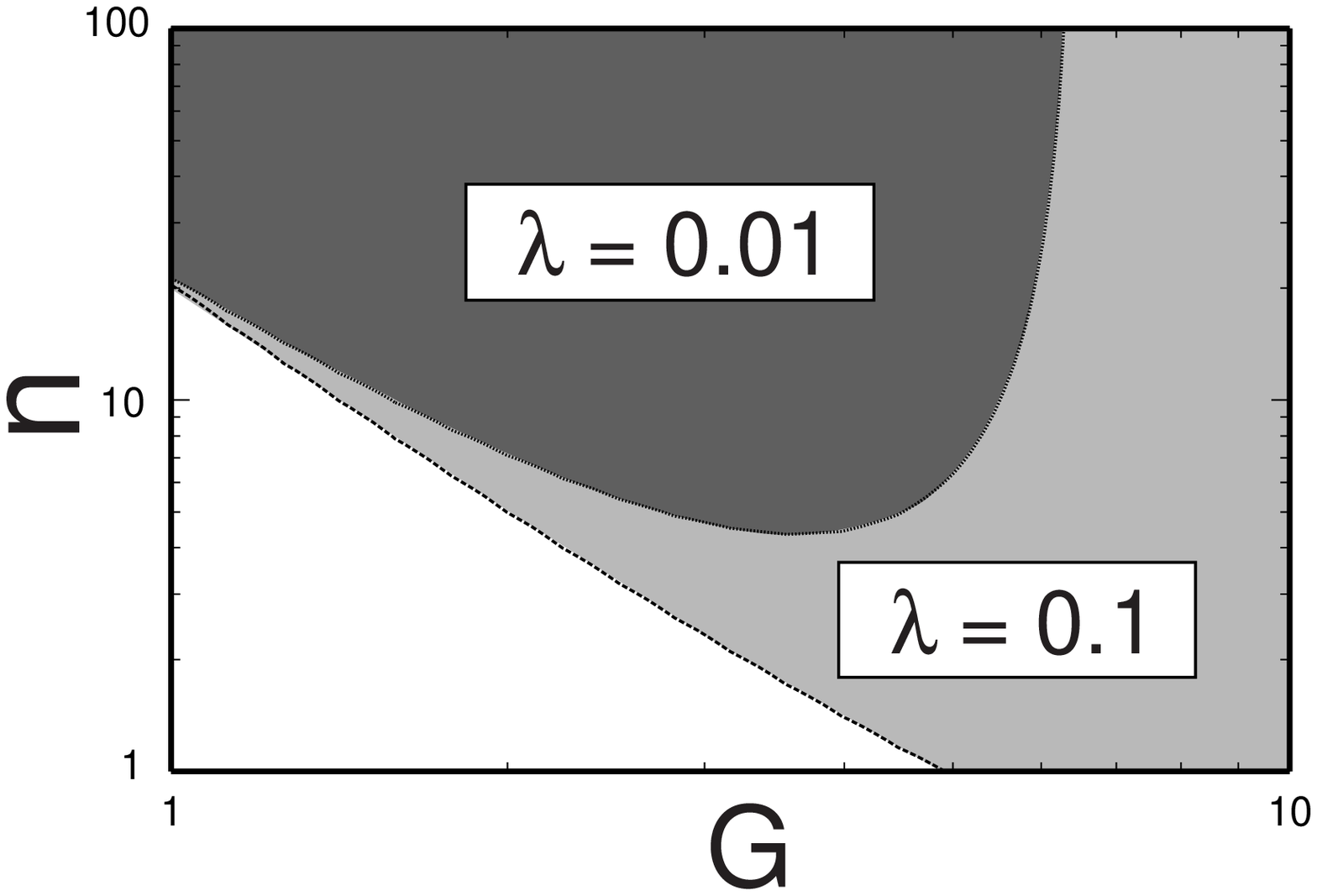}
\end{minipage}
\caption{The allowed region (shadowed) of the scalar self-coupling constant $\lambda$ (left) and the Yukawa coupling constant $G$ (right) for the Fermi ball to be stable. The assumptions are the same as those in Figure \ref{onphase.fig}. We see that the allowed regions broaden as $n$ increases. \label{glfixphase.fig}}
\end{figure}

Let us consider the simplified model to examine how the number of the
fermion flavors $n$ affects the stability of the Fermi ball in case of
$N_i^{(a)}=c^{(a)}N_i$. We assume that $\Psi_i$ belongs to a multiplet
of the internal symmetry with a common Yukawa coupling constant $G$ and
also assume that the fermion number is common to the flavor, i.e.,
$N_i=N$. Under these assumptions, the coefficient ${\cal C}$ is
independent of $N$ and dependent on $\lambda$, $G$ and $n$ from
Eq.(\ref{C.eqn2}). We evaluate Eq.(\ref{C.eqn2}) using a numerical
integration, and obtain the stable region of the parameters where ${\cal
C}$ is positive (see Figures \ref{onphase.fig} and
\ref{glfixphase.fig}).
These figures show that the allowed regions of the parameters exist for the Fermi ball to be stable against the fragmentation (the shadowed regions in the figures). We see in the figures that the allowed region broadens as the number of the flavors $n$ increases.

%%%%%%%%%%
%%%%%%%%%%
\section{CONCLUSION}
\label{concl.sec}
%%%%%%%%%%
%%%%%%%%%%
We have considered a model for the Fermi ball in which the fermions with
multi-flavors $\Psi_i$ ($1\leq i\leq n$) are coupled to the scalar field
$\phi$ and the total fermion number of each $i$-th flavor is fixed as
$N_i$. We have examined the region of the parameters for the Fermi ball
to be stable against fragmentation, and how the number of the
fermion flavors $n$ affects the stability.

We have considered the thin-wall Fermi ball, i.e., the radius $R$ is
much larger than the wall thickness $\delta_b$. We have taken into
account the effect due to the finite wall thickness by the perturbation
expansion with respect to $\delta_b/R$. In the leading order thin-wall
approximation, we have compared the energy of the initial state of a
single Fermi ball and that of the final state of fragmented $m$ Fermi
balls, with the total fermion number $N_i$ of each flavor $i$ being
conserved, $\sum_{a=1}^{m}N_i^{(a)}=N_i$. We have found that the former
is smaller than the latter and thus the Fermi ball is stable against
fragmentation, except for the special case of $N_i^{(a)}=c^{(a)}N_i$
with $\sum_{a=1}^{m}c^{(a)}=1$. This situation that the Fermi ball is
stable in most cases is characteristic of the case with multi-flavor of
fermions, and qualitatively different from the case of a single
flavor. In the special case of $N_i^{(a)}=c^{(a)}N_i$, the two states
have the same energy in the leading order approximation and the
next-to-leading order correction term $\delta E$ determines the
stability. There we have found that the energy of the initial state is
$E_0 +{\cal C}v$ and that of the fragmented states is $E_0 +m{\cal C}v$,
where $v$ is a symmetry breaking scale and ${\cal C}$ is a coefficient
dependent on the scalar self-coupling constant $\lambda$, the Yukawa
coupling constant $G_i$ and the fermion number $N_i$. This tells us that
even in that case the Fermi ball is stable when ${\cal C}$ takes a
positive value in the parameter region of $\lambda$, $G_i$ and $N_i$.

We have considered the simplified model in which a multiplet of fermions
has a common $G_i$ and a common $N_i$ for each flavor $i$. We have found
that the allowed region of the parameters for the Fermi ball to be
stable exists and broadens as the multiplet dimension $n$ increases.

%%%%%%%%%%
%%%%%%%%%%


\begin{thebibliography}{999}
%%%%%%%%%%
%%%%%%%%%%
\bibitem{MAC}A. L. Macpherson and B. A. Campbell, Phys. Lett. {\bf B347}
(1995) 205.
%%%
\bibitem{MOR}J. R. Morris, Phys. Rev. {\bf D59} (1999) 023513.
%%%
\bibitem{LEE}T. D. Lee and Y. Pang, Phys. Rep. {\bf 221} (1992) 251.
%%%
\bibitem{DVA} For the examples of zero-mode fermions bound in the domain wall, see:  G. Dvali and M. Shifman, Nucl. Phys. {\bf B504} (1997) 127 ; M. Sakamoto and M. Tachibana, Phys. Lett. {\bf B458} (1999) 231.
%%%
\bibitem{ARA}J. Arafune, T. Yoshida, S. Nakamura and K. Ogure,
Phys. Rev. {\bf D62} (2000) 105013.
%%%
\bibitem{YOS}T. Yoshida, K. Ogure and J. Arafune, hep-ph/0210062, to appear in Phys. Rev. {\bf D}.
%%%
\bibitem{OGU}K. Ogure, T. Yoshida and J. Arafune, hep-ph/0212332.
%%%
\end{thebibliography}
\end{document}